\documentclass[11pt,a4paper,nofootinbib,showpacs]{article}
\pdfoutput=1

\usepackage{jcappub}

\usepackage[utf8]{inputenc}
\usepackage[english]{babel}
\usepackage[T1]{fontenc}
\usepackage{graphicx}
\usepackage{amsmath}
\usepackage{subfig}
\usepackage{hyperref}
\usepackage{color}
\usepackage{amssymb}
\usepackage[font=scriptsize,labelfont=bf]{caption}
\usepackage{appendix}
\usepackage{slashed}
\usepackage{slantsc}
\usepackage{braket}
\usepackage{mathrsfs}
\usepackage{subfig}
\usepackage{placeins}
\usepackage{fullpage}
\usepackage{sidecap}
\usepackage{float}
\usepackage{wrapfig}
\usepackage[dvipsnames]{xcolor}
\usepackage{dsfont}

\usepackage{times}
\usepackage{natbib}
\usepackage{epsfig}
\usepackage{bm}
\usepackage{amsmath}
\usepackage{amssymb}

\newcommand{\aj}{Astron. J.}
\newcommand{\aap}{Astron. Astrophys.}
\newcommand{\cqg}{Class. Quant. Grav.}
\newcommand{\mnras}{Mon. Not. R. Astron. Soc.}

\newcommand{\jcap}{J. Cosmol. Astropart. Phys.}

\newcommand{\grg}{Gen. Relativ. Gravit.}
\newcommand{\apj}{Astrophys. J.}
\newcommand{\prd}{Phys. Rev. D}

\newcommand{\up}[1]{{\rm #1}}

\newcommand{\drr}{\delta r}
\newcommand{\kcut}{k_\up{IR}}
\newcommand{\rbar}{\bar r}
\newcommand{\dDI}{\delta\mathcal{D}^{\dT_o}_L}
\newcommand{\dDII}{\delta\mathcal{D}^\up{std}_L}
\newcommand{\beeq}{\begin{equation}}
\newcommand{\eneq}{\end{equation}}
\newcommand{\bear}{\begin{eqnarray}}
\newcommand{\enar}{\end{eqnarray}}
\newcommand{\RA}{\rightarrow}
\newcommand{\HH}{\mathcal{H}}   
\newcommand{\gbar}{\bar g}      
\newcommand{\dz}{\delta z}           
      
\newcommand{\ddL}{\delta\mathcal{D}_L} 
\newcommand{\UU}{u}         
\newcommand{\dT}{\delta\tau}
\definecolor{red}{rgb}{0.00,0.00,0.00}

\begin{document}

\title{Gauge-Invariance and Infrared Divergences in the Luminosity Distance}

\author[a,c]{Sang Gyu Biern}
\author[a,b]{and Jaiyul Yoo}
\affiliation[a]{Center for Theoretical Astrophysics and Cosmology,
Institute for Computational Science, University of Z\"urich,
Winterthurerstrasse 190, CH-8057, Z\"urich, Switzerland}
\affiliation[b]{Physics Institute, University of Z\"urich,
Winterthurerstrasse 190, CH-8057, Z\"urich, Switzerland}
\affiliation[c]{Asia Pacific Center for Theoretical Physics, Pohang, 790-784, Korea}
\emailAdd{sgbiern@physik.uzh.ch}
\emailAdd{jyoo@physik.uzh.ch}
\date{\today}

\abstract{
Measurements of the luminosity distance have played a key role
in discovering the late-time cosmic acceleration. 
However, when accounting for inhomogeneities in the Universe, its 
interpretation has been plagued with 
infrared divergences in its theoretical predictions, which are in some 
cases used to explain 
the cosmic acceleration without dark energy. The infrared divergences in most 
calculations are artificially removed by imposing an infrared cut-off scale.
We show that a gauge-invariant calculation of the luminosity
distance is devoid of such divergences and consistent with the equivalence 
principle, eliminating the need to impose a cut-off scale.
We present proper numerical calculations of the luminosity distance using the 
gauge-invariant expression
and demonstrate that the numerical results with an {\it ad hoc} 
cut-off scale
in previous calculations have negligible systematic errors as long as
the cut-off scale is larger than the horizon scale. We discuss the origin
of infrared divergences and their cancellation in the luminosity distance.
}


\arxivnumber{1606.01910}

\maketitle
\flushbottom

\section{Introduction}

The late-time acceleration of the Universe was convincingly demonstrated
by measuring the luminosity distances of distant supernovas, establishing the
existence of exotic energy component of the Universe, or dark energy
\cite{PEADET99,RIFIET98}.
Supporting evidence for dark energy accumulated over time through other 
measurements such as the microwave background anisotropies 
and the acoustic peak position in galaxy clustering (e.g., \cite{PLANCK13,
EIZEET05}). Despite these recent developments,
the measurements of the luminosity distance provide 
the most critical and direct evidence for the late-time cosmic
acceleration.

In search of the best set of cosmological parameters,
the luminosity distance measurements are often compared to the theoretical
predictions in a homogeneous universe. This procedure ignores
the effects of inhomogeneities in our Universe on the luminosity distance.
The utmost relevant question in this case
is whether these effects are significant enough
to change the outcome in the conventional approach, compared to the measurement
precision. A significant amount of efforts are made to quantify the contribution of inhomogeneities~\cite{BEDUET14,HUGR06,BODUGA06}, and the consensus is that while such effects are important, no dramatic change arises in its theoretical interpretation.

However, there has been a claim \cite{BAMARI05}
that the luminosity distance measurements can
be fully explained by the statistical fluctuations of metric perturbations in our Universe
without invoking the existence of dark energy. In fact, the relativistic metric perturbations
on super-horizon scales are spatially constant, albeit varying in time,
categorically affecting the luminosity distance measurements in our
local Universe (or the Universe within our horizon). Such perturbations 
on super-horizon scales are generated during the inflationary period at the
early time, providing the physical mechanism for the observed anisotropies
in the microwave background radiation. If the super-horizon perturbations such as the curvature perturbation $\zeta$ are generated to a very large scale equivalent to $\sim500$ 
$e$-folding of the current horizon scale, 
the variance of the luminosity distance reaches
an order unity and we may find ourselves in a universe with such fluctuations
but without dark energy, claims \cite{BAMARI05} (see also \cite{KOMAET05a}
for a similar argument based on the averaged expansion).

This idea has been dismissed in the community, simply because of the largest scale or 
its insensitivity to the cut-off scale~$\kcut$ in the infrared --- The variance
of the luminosity distance diverges logarithmically with~$\kcut$ if the curvature spectrum is scale-invariant (faster if red tilted). Therefore,
spelled explicitly or not, all the practitioners 
impose an {\it ad hoc} scale around 
$\kcut\simeq\HH_o$ in their computation of the luminosity distance, 
where $\HH_o$ is the conformal Hubble parameter today. 
However, the introduction of such ad hoc scale is unsatisfactory, and more
importantly the Universe is likely to be more inhomogeneous or chaotic 
beyond the largest scale~$\kcut$ with inflationary fluctuations;
it would require a fine-tuning to assume the absence of 
inhomogeneity below the level of inflationary curvature 
perturbation~$(\sim10^{-5})$ beyond~$\kcut$.

In this paper, we provide a complete resolution of this issue 
by proving that there are {\it no} infrared divergences in the luminosity
distance and the fluctuations on super-horizon scales indeed {\it cancel}.
The physical explanation for such cancellation is the equivalence principle,
according to which the super-horizon modes of gravity {\it cannot}
affect the small-scale dynamics such as the luminosity distance measurements.
While the correct expression for the luminosity distance is gauge-invariant, the calculations of the luminosity distance in literature
were in most cases 
based on incorrect {\it gauge-dependent} expressions even at the linear order,
yielding non-physical features including infrared divergences.
We demonstrate how such errors are made and restoring the gauge invariance can cure such pathology in the theoretical prediction.

This paper is organized as follows: In section \ref{Sec:GI}, we show the gauge-invariance of the luminosity distance fluctuation and present the standard (incorrect) expression frequently used in the previous research.  In section \ref{Sec:dtau}, we review the coordinate lapse at the observation, which is a key missing part of the full (correct) luminosity distance expression.  In section \ref{Sec:IR}, we show how the infrared divergence arises in the standard expression of the luminosity distance variance and how the divergence disappears by considering the residual contribution. In section \ref{Sec:cutoff}, we investigate the significance of the super-horizon perturbation contribution to the luminosity distance variance. The complete variances of each contribution of the luminosity distance are presented in~\ref{Sec:var}. In section \ref{Sec:discussion}, we discuss our results. The details of the derivation of the luminosity distance in an inhomogeneous universe are presented in appendix. We show the expression of the wave vector distortion with perturbation variables in appendix~\ref{App:wavevector}, including a discussion of the wave vector perturbation at the observation. In appendix~\ref{App:distortion}, the detailed expressions of the luminosity distance components without choosing any gauge. The perturbation solutions in the flat $\Lambda$CDM universe is presented in appendix~\ref{App:sol}. Finally, in appendix~\ref{App:EP}, we verify the validity of the computation of the luminosity distance by checking that it is consistent with the equivalence principle.

\section{Gauge-invariance of the Luminosity distance in an inhomogeneous universe}
\label{Sec:GI}
The luminosity distance in the background is simply
$\bar{\mathcal{D}}_L(z)=\rbar_z/(1+z)$, where~$z$ is the redshift parameter,
the comoving radial distance to the redshift is
$\rbar_z=\int_0^z dz/H(z)$, and the Hubble parameter is~$H(z)$.
In the presence of inhomogeneities in our 
Universe, we parametrize the metric perturbations~$\delta g_{\mu\nu}$ in 
the Friedmann-Robertson-Walker (FRW) universe with four scalar perturbations
($\alpha,\beta,\varphi,\gamma$) as
\beeq
\delta g_{00}=-2a^2\alpha~,\quad \delta g_{0i}=-a^2\beta_{,i}~,\quad
\delta g_{ij}=2a^2\left(\varphi\gbar_{ij}+\gamma_{,i|j}\right),
\eneq
where $a$ is the scale factor, $\gbar_{ij}$ is the background 3-metric,
and commas represent the ordinary derivative while the vertical bar represents
the covariant derivative with~$\gbar_{ij}$.
We assumed a flat universe without vector or tensor perturbations
for simplicity. Popular choices of gauge conditions in literature are
the conformal Newtonian gauge ($\beta=0=\gamma$), the comoving gauge ($\UU_i\equiv
av_{,i}=0=\gamma$),
and the synchronous gauge ($\alpha=0=\beta$), where
$\UU^i$ is the spatial component of the observer four velocity.
In a flat $\Lambda$CDM universe the comoving gauge condition yields $\alpha=0$ (hence comoving-synchronous gauge), while the synchronous gauge condition alone has a residual gauge mode carried by~$\gamma$ (see, e.g., \cite{YOO14b}).

Using the geometric approach \cite{YOO14a},
the dimensionless fluctuation $\delta\mathcal D_L$ in the luminosity distance can be expressed as a function of the \emph{observed} redshift $z$ and angle $\hat{\bm n}$ (see  also \cite{SASAK87} for a complete derivation):
\bear
\label{dL}
\mathcal D_L\equiv\bar{\mathcal D}_L\left(1+\delta\mathcal D_L\right),~~~
\ddL(z,\hat{\bm n})   
&=&\dz+{\drr\over\rbar_z}-\kappa+\Xi~,
\end{eqnarray}
where the distortion~$\dz$ in the observed redshift 
$1+z\equiv(1+\dz)/a_s$ of the source is
\bear
\label{dz}
\dz&=&~\mathcal H_o\delta\tau_o+\left[-\varphi-\mathcal I+ \mathcal V_\| \right]^s_o -\int_0^{\bar r_z}d\bar r~\mathcal I'\, ,
\enar
the
radial distortion~$\drr$ and the angular distortion$~\kappa$
in the comoving source position in spherical coordinates are
\bear
\label{dr}
\drr&=&~\delta\tau_o-\frac{1}{\mathcal H_z}\delta z +\left[ -\beta-\gamma'-\partial_\|\gamma	\right]^s_o +\int_0^{\bar r_z}d\bar r~\mathcal I\, ,\\
\label{kappa}
\kappa&=&-\mathcal V_{\|o} +\frac{1}{\bar r_z}\partial_\|\gamma_o +\frac{1}{2}\Delta_\perp \gamma_s-\frac{1}{2}\left[\mathcal I\right]^s_o -\int_0^{\bar r_z}d\bar r \left\{\frac{\bar r}{\bar r_z}\mathcal I'+\frac{1}{2}\frac{\left(\bar r_z-\bar r\right)\bar r}{\bar r_z}\left(\mathcal I''-\bar \Delta\mathcal I\right)
	\right\}\, ,
\enar
and the frame distortion $\Xi$ at the source (see appendix~\ref{App:distortion} for the derivation) is
\begin{equation}
\label{eq:frame}
\Xi=\varphi_s+\frac{1}{2}\Delta_\perp \gamma_s+\frac{1}{\bar r_z}\partial_\|\gamma_s\, .
\end{equation}
The notation convention is that primes represent
the derivative with respect to the conformal time~$\tau$, 
$\Delta$ ($\Delta_\perp$) is the
Laplacian operator in spatial three-dimension (two-dimensional spheres),\footnote{$\bar\Delta$ in the integration of $\kappa$ indicates the three-dimensional Laplacian operator in the bar coordinate $\bar r\bm{\hat n}$.} and $\partial_\|$ is a spatial derivative along the line-of-sight. The suffix $o$ ($s$) indicates the observation (source) position, and $\dT_o$ is the coordinate
lapse at the observer position (see Eq. (\ref{dTo})), we defined two gauge-invariant variables $\mathcal V_\|\equiv \partial_\|(-v+\beta+\gamma')$ and $\mathcal I\equiv\alpha-\beta'-\varphi-\gamma''$. The rectangular bracket $[\cdots]^s_o$ indicates a field difference between the source and the observation i.e., $[X]^s_o = X_s-X_o$. It is noted that the angular distortion $\kappa$ has the radial velocity component~$\mathcal V_\|$
at the observer position.   Under the temporal coordinate transformation $\tau\RA\tau+T(x^a)$, each component of~$\ddL$ transforms as $\delta\tau_o\RA\delta\tau_o+T_o$, $\varphi\RA\varphi-\HH T$, $\beta+\gamma'\RA\beta+\gamma'-T$ and $\dz\RA\dz+\HH_s T_s$, while $\drr$ and $\kappa$ remain invariant under
the temporal gauge transformation. Temporal and spatial gauge modes are completely canceled in $\delta\mathcal D_L$: $\delta\mathcal D_L$ is gauge-invariant.

The full expression of~$\ddL$ was first presented with a general metric
representation in \cite{SASAK87}
with an explicit check of its gauge-invariance, though no numerical computation
of the variance was made.
Despite the statements claimed in many recent works
\cite{BAMARI05,BODUGA06,BEGAET12a,BEMAET12a,BEGAET13,FAGAET13,
UMCLMA14a,BEDUET14},
only a few \cite{SASAK87,YOO10,YOO14a} in literature actually
demonstrated the gauge-invariance of~$\ddL$.
The computation of the luminosity distance is mostly
performed in the conformal Newtonian gauge, while
a few \cite{BAMARI05,JESCHI12,FAGAET13} in the synchronous gauge.
As the equation for~$\ddL$ is gauge-invariant, it is just a matter of
convenience which gauge condition is chosen for the computation of~$\ddL$,
and the computations in any gauge conditions should agree. Unfortunately, when a certain gauge condition is adopted, an error is often made and the resulting expressions often used in literature are \emph{not} gauge invariant.

To elaborate on this statement, we split the full (gauge-invariant) expression of~$\ddL$ into two 
components and adopt the conformal Newtonian gauge condition ($\alpha=-\varphi\equiv\Psi$, $\beta=0=\gamma$, and $\mathcal V_\| \equiv V_\|$):
\bear
\label{Eq:dDL}
\delta\mathcal D_L &\equiv & \delta\mathcal D_L^\text{std}+\delta\mathcal D_L^{\delta\tau_o}\, ,\\
\label{Eq:dDL_dtau}
\dDI&=&\left(\HH_o+\frac{1}{\rbar_z}-\frac{\HH_o}{\HH_z\rbar_z}\right)\dT_o\, ,
\enar
where the suffix $z$ indicates that quantities, such as $\bar r_z$ and $\mathcal H_z$, are evaluated at redshift $z$. The first component~$\dDII$ is the 
expression for the luminosity distance widely used in literature 
\cite{BODUGA06,BEGAET12a,BEMAET12a,BEGAET13,FAGAET13,UMCLMA14a,BEDUET14},
while the other component~$\dDI$ is the residual part missing in $\delta\mathcal D_L^\text{std}$. The sum of the two is the full expression $\delta\mathcal D_L$ and is gauge-invariant, while $\delta\mathcal D_L^\text{std}$ and $\delta\mathcal D_L^{\delta\tau_o}$ are not separately gauge-invariant. Since $\dT_o\RA\dT_o+T_o$,
it is evident that $\delta\mathcal D_L^{\delta\tau_o}$ is not gauge-invariant and using~$\dDII$ in place of~$\ddL$
for the luminosity distance  also breaks the gauge-invariance. This incorrect use of $\delta\mathcal D_L^\text{std}$ for $\delta\mathcal D_L$ is the reason the computations of the luminosity distance in the 
conformal Newtonian gauge led to {\it inconsistent} results. We note that the correct gauge-invariant calculation $\delta\mathcal D_L= \delta\mathcal D_L^\text{std}+\delta\mathcal D_L^{\delta\tau_o}$ has been presented~\cite{YOO14a,SASAK87,YOSC16,JESC14} with the coordinate lapse $\delta\tau_o$ properly considered.

From now on, we choose the conformal Newtonian gauge, and decompose the standard expression $\delta\mathcal D_L^\text{std}$ into the velocity $\delta\mathcal D_L^V$, the gravitational potential $\delta \mathcal D_L^\Psi$, and the lensing $\delta\mathcal D_L^\text{lens.}$ contributions i.e., $\delta\mathcal D_L^\text{std}=\delta \mathcal D_L^V+\delta\mathcal D_L^\Psi+\delta\mathcal D_L^\text{lens.}$:
\begin{align}
\label{Eq:components}
	\delta \mathcal D_L^V =&~\left(1-\frac{1}{\mathcal H_z\bar r_z}\right)V_{\|s} +\frac{1}{\mathcal H_z\bar r_z}V_{\|o}\, ,~~~~~~~~~~~~ \\
	\delta \mathcal D_L^\text{lens.} =&-\int_0^{\bar r_z}d\bar r \left\{(\bar r_z-\bar r)\frac{\bar r}{\bar r_z}\bar\Delta\Psi \right\}\, ,\nonumber\\ 
	\delta \mathcal D_L^{\Psi} =&~\frac{1}{\mathcal H_z\bar r_z}\Psi_s-\frac{1}{\mathcal H_z\bar r_z}\Psi_o-\Psi_s+\int_0^{\bar r_z}d\bar r \left\{ \frac{2}{\bar r_z}\Psi +2\left( \frac{1}{\mathcal H_z\bar r_z}+\frac{\bar r}{\bar r_z}-1\right)\Psi'+\left(\bar r_z-\bar r\right)\frac{\bar r}{\bar r_z}\Psi''	\right\}\, .\nonumber
\end{align}
 Note that the velocity and the lensing contributions are composed with respectively single ($V_\| \propto \partial_\| \Psi$) and double ($\Delta\Psi$) spatial partial derivative of the gravitational potential $\Psi$, and the gravitational potential contribution is devoid of any spatial differentiation of $\Psi$.

\section{Coordinate lapse at the observation}
\label{Sec:dtau}

The (conformal-) coordinate lapse $\delta\tau_o$ at the observation  is the key component, and it is readily computed in~\cite{YOO14b} (see also \cite{JESC14}). In this section we review how $\delta\tau_o$ arises in an inhomogeneous universe. Physically, the coordinate lapse at the observation originates from the fact that the (physical-) coordinate time $t_o$ at the observation  (or the conformal-coordinate time $\tau_o$) does not correspond to the proper time $\mathcal T_o$  (or the coordinate time in the synchronous or the comoving gauges) of the observer. 

The physical-coordinate time $t$ and the proper-time $\mathcal T$ of the observer  are related to the time component of the four-velocity $u^\mu$ of the observer as
\begin{equation}
\label{Eq:dtdT}
	u^t =\frac{dt}{d\mathcal T }= 1-\alpha\, .
\end{equation}
For fixed proper time $\mathcal T_o$ measured by the observer, the corresponding coordinate time $t_o$ is obtained by integrating (\ref{Eq:dtdT}) along the path of the observer:
\begin{equation}
\label{Eq:to}
	t_o=t(\mathcal T_o,\bm x)= \int_0^{\mathcal T_o} \left(1-\alpha(\mathcal T,\bm x)\right)d\mathcal T= \mathcal T_o - \int_0^{\mathcal T_o}\alpha(\mathcal T,\bm x) d\mathcal T\equiv\mathcal T_o+\delta t_o(\mathcal T_o,\bm x)\, ,
\end{equation}
where the difference between $t_o$ and $\mathcal T_o$ is the (physical-) coordinate lapse $\delta t_o$ at the observation. It is noted that for a given proper time $\mathcal T_o$ different observers at different spatial postions find themselves at different time coordinates $t_o$.  

In the conformal-coordinate, the conformal-coordinate time $\tau_o$  at the observation is decomposed as $\tau_o=\bar\tau_o+\delta\tau_o$, where $\bar\tau_o$ and $\delta\tau_o$  correspond to respectively the conformally transformed proper time and the (conformal-) coordinate lapse at the observation. From the conformal transformation, $\mathcal T_o$ is related to $\bar\tau_o$ as $\mathcal T_o = \int_{0}^{\bar\tau_o}a(\tau)d\tau$, and $\bar\tau_o$ is \emph{uniquely} determined as
\begin{equation}
	\bar\tau_o = \int_0^\infty \frac{dz'}{H(z')}\, .
\end{equation}
In addition, the conformal-coordinate lapse $\delta\tau_o$ at the observation  is derived from the conformal transformation $\delta t_o=a(\bar\tau_o)\delta\tau_o$:
\begin{equation}
\label{dTo}
\delta\tau_o = -\frac{1}{a(\bar\tau_o)}\int_0^{\bar\tau_o}\alpha(\bar\tau,\bm 0) a(\bar\tau)d\bar\tau\, .
\end{equation}
Note that $\alpha(\bar\tau,\bm 0)$ indicates that we set the spatial position of the observer as $\bm x_o = \bm 0$. 

In the synchronous and the comoving gauges, $\delta t_o$ vanishes because of $\alpha=0$: the (physical-) coordinate time in these gauges is sychronized to the proper time of the observer. However, in the conformal Newtonian gauge, where many previous studies~(e.g., \cite{HUGR06,BACBM14,BODUGA06,BEGAET12a,BEMAET12a,BEGAET13,FAGAET13, UMCLMA14a,BEDUET14}) were performed, the metric perturbation $\alpha$ corresponds to the gravitational potential $\Psi$, and $\delta t_o$ can be understood as the gravitational time dilation. The (physical-) coordinate time therefore is not necessarily equivalent to the proper-time of the observer. Though $a(\bar\tau_o)$ is different from $a(\tau_o)$ in the conformal Newtonian gauge i.e., $a(\tau_o)=a(\bar\tau_o)(1+\mathcal H_o\delta\tau_o)$, this subtle difference is often neglected in the previous studies.

Neglecting $\delta\tau_o$ in the luminosity distance yields two theoretical problems. The first problem is that the luminosity distance is no longer gauge-invariant without $\delta\tau_o$, as shown in the previous section. The second problem is that the luminosity distance without $\delta\tau_o$ is inconsistent with the equivalence principle (see appendix~\ref{App:EP}), and this inconsistency leads to the infrared divergence in the luminosity distance variance.  In the following section, we will discuss how the problematic $\delta\tau_o$ is responsible for the infrared divergence in the luminosity distance and how this problem is resolved by properly accounting for the coordinate lapse.

\section{Infrared divergences in the luminosity distance}
\label{Sec:IR}

In section~\ref{Sec:GI}, the standard expression $\delta\mathcal D_L^\text{std}$ of the luminosity distance and the residual expression $\delta\mathcal D_L^{\delta\tau_o}$ are presented. In this section, we show how the infrared divergence arises in $\delta\mathcal D_L^{\delta\tau_o}$ and present numerical computations of $\delta\mathcal D_L^\text{std}$ and $\delta\mathcal D_L^{\text{std}}+\delta\mathcal D_L^{\delta\tau_o}$.

 In the conformal Newtonian gauge, where the gravitational potential $\Psi$ is related to the velocity potential $\Phi_v$ as $\Psi = -\frac{1}{a}(a\Phi_v)'$ (see appendix~\ref{App:sol}), the coordinate lapse at the observation $\delta\tau_o$ is determined as
\begin{equation}
	\delta\tau_o = -\frac{1}{a(\bar\tau_o)} \int_0^{\bar\tau_o}a(\bar\tau)\Psi(\bar\tau,\bm 0)d\bar\tau =  \Phi_{vo}\, .
\end{equation}
The peculiar velocity $\bm V$ is generated by the velocity potential as $\bm V = \bm\nabla \Phi_v$, and the velocity potential can be written as $\Phi_v(\tau,\bm x)=D_V(\tau)\zeta(\bm x)$, where $D_V(\tau)$ and $\zeta$ are repectively the velocity growth function and the curvature perturbation in the comoving gauge, as shown in appendix \ref{App:sol}. The variance of $\delta\tau_o$ is
\begin{equation}
\label{Eq:Var_dtau}
	\left<\delta\tau_o^2\right> = D_V^2(\bar\tau_o) \left<\zeta_o^2\right> = D_V^2(\bar\tau_o) \lim_{\varepsilon\rightarrow0}\int_\varepsilon^\infty \frac{dk}{k}\Delta^2_\zeta(k)\, ,
\end{equation}
where $\Delta_\zeta^2(k)$ is the dimensionless power spectrum of the curvature perturbation, and  $\Delta_\zeta^2(k)=A_s(k/k_o)^{n_s-1}$ for $k\ll k_\text{eq}$, where $n_s$ is the scalar spectral index, $k_o$ is the pivot wave number ($k_o=0.05~\text{Mpc}^{-1}$), $A_s$ is the scalar amplitude,  and $k_\text{eq}$ is the scale corresponding to matter-radiation equality. Solving  Eq. (\ref{Eq:Var_dtau}) further, we can isolate the divergent and the convergent parts as
\begin{equation}
\label{Eq:var_dtau2}
	\left<\delta\tau_o^2\right> = D_V^2(\bar\tau_o) \left\{\lim_{\varepsilon\rightarrow0}\frac{ \varepsilon ^{n_s-1}}{1-n_s} - \frac{k_\text{IR}^{n_s-1}}{1-n_s} + \int_{k_\text{IR}}^{k_\text{UV}} \frac{dk}{k}\Delta_\zeta^2(k)\right\}\, ,
\end{equation}
where $k_\text{IR}$ and $k_\text{UV}$ are the infrared and the ultraviolet cutoffs, respectively, ($k_\text{IR}\ll k_\text{eq}$). Note that the first term in the bracket diverges when $n_s\leqq  1$. Therefore, we obtain a large variance due to the curvature perturbation $\zeta$ on large scales $k< k_\text{IR}$.  In other words, the long-wavelength  mode perturbations are much more important than the other modes in the variance of $\delta\tau_o$. As shown in appendix \ref{App:sol}, the gravitational potential is also expressed with the curvature perturbation and its growth function as $\Psi=D_\Psi\zeta$, and it also leads to the same infrared divergence. However, the velocity and the lensing contributions in Eq. (\ref{Eq:components}) do not yield the infrared divergence since they depend on $\partial_\|\Psi$ and $\Delta\Psi$, each of which is multiplied by $k$ and $k^2$, respectively.

\begin{figure}[t]
\centering
\includegraphics[scale=1.0]{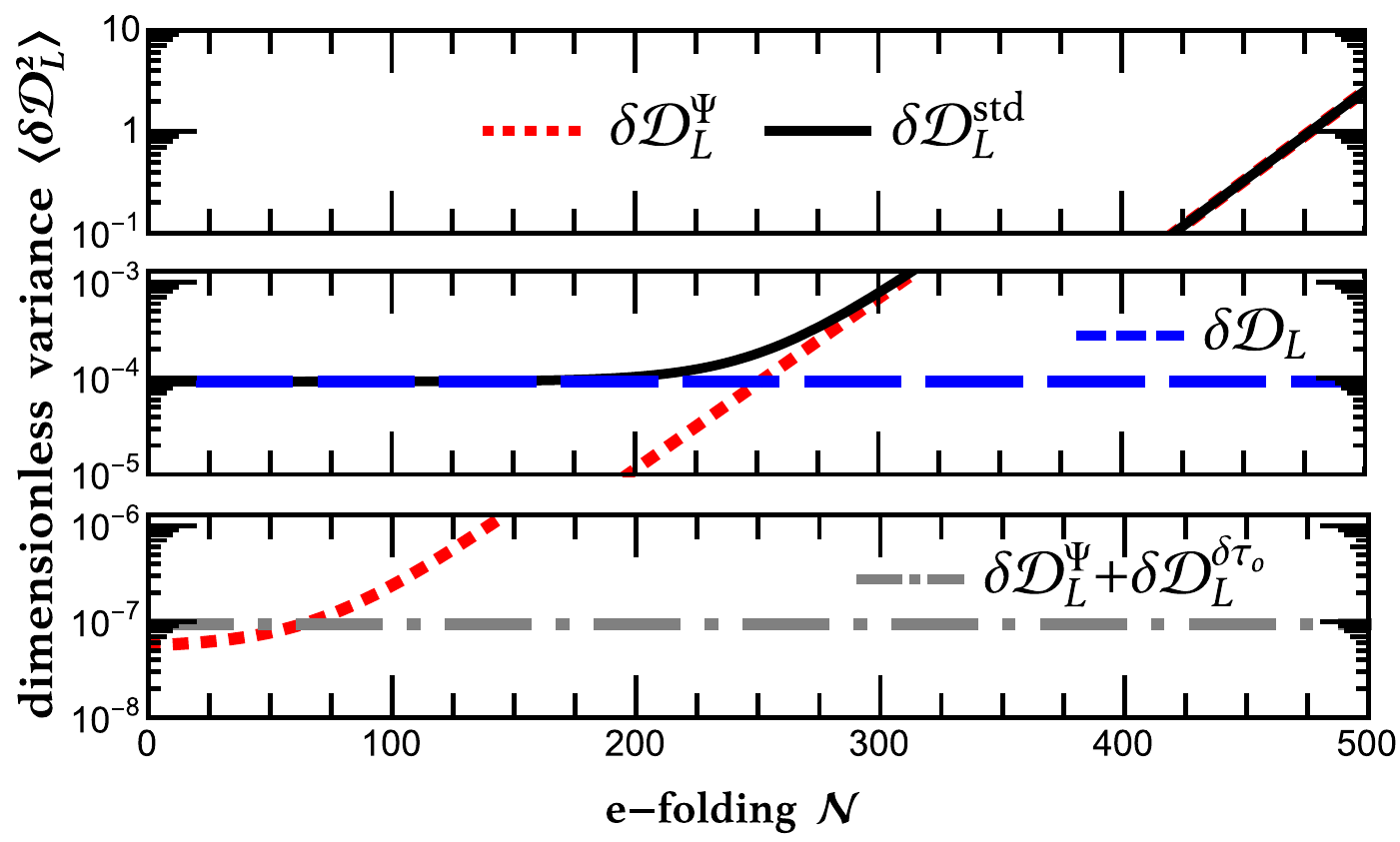}
\caption{Variance $\left\langle\ddL^2\right\rangle$ of the luminosity distance
fluctuation as a function of infrared cut-off scale 
($\kcut\equiv\HH_o\exp[-N]$) at $z=0.1$, where $N$ approximately corresponds 
to the number of $e$-folding of the largest scale 
with curvature perturbations generated during the inflationary epoch.
The standard expression~$\dDII$ (solid) often used in literature is 
{\it gauge-dependent}, diverging with~$N$, while the correct
{\it gauge-invariant} calculation~$\ddL=\dDII+\dDI$ (dashed)
is indeed finite
with long-modes of gravity cancelled. For numerical calculations, we assume
a flat $\Lambda$CDM universe with the spectral index $n_s=0.96$,
the matter density $\Omega_m=0.3$, the scalar amplitude $A_s=2\times10^{-9}$, 
and the Hubble parameter $h=0.68$. We set the upper limit of momentum integration $k_\text{max}=300\mathcal H_o$ for numerical integrations.}
\label{IR}
\end{figure}

Now let us study these contributions to the luminosity distance variance, accounting for fixed $k_\text{UV}$ and varying $k_\text{IR}$. Figure~\ref{IR} illustrates how the variance of the standard expression $\dDII$ at $z=0.1$ varies with respect to the e-folding $\mathcal N$, where $e^{-\mathcal N}\equiv k_\text{IR}/\mathcal H_o$. As first noted in \cite{BAMARI05}, the variance of~$\dDII$ increases exponentially with infrared cutoff scale~$\kcut$ (or larger $\mathcal N$), and if the scale is sufficiently large, the variance can reach an order unity  to forgo the need for dark energy in explaining the luminosity distance  measurements. This dramatic increase in the variance of~$\dDII$ is due to the infrared divergence  of the contribution of the gravitational potential $\delta\mathcal D_L^\Psi$ (dotted) on super-horizon scales --- logarithmically for the scale-invariant power spectrum ($n_s=1$) and
faster for the power spectrum with red tilt ($n_s<1$). We emphasize that $\dDII$ is {\it gauge-dependent} and is {\it not} a correct expression for the luminosity distance~$\ddL$ in an inhomogeneous universe, and this divergent behavior of $\dDII$ is clearly cancelled with the variance of $\delta\mathcal D_L^{\delta\tau_o}$. As a result, the variance computed by using the correct expression~$\ddL$ (dashed curve) at $z=0.1$ changes little with respect to the super-horizon scale cutoff.

While long-mode fluctuations of gravity can modulate the small-scale dynamics,
the impact of such fluctuations should decrease as the scale of the
fluctuations increases. The equivalence principle implies that fluctuations 
on super-horizon scales affect everything altogether and there is {\it no}
way to tell the existence of uniform gravitational fields, in direct
conflict with the calculation of~$\dDII$ (solid) in Figure~\ref{IR}.
This infrared-divergence of the gravitational potential~$\ddL^\Psi$
(dotted)
is cancelled (dot-dashed) by the missing contribution~$\dDI$ in the
luminosity distance,
and the variance of the luminosity distance is indeed finite
and devoid of such divergences, shown as the dashed curve.

\section{Super-horizon perturbation contribution to the luminosity distance variance}
 \label{Sec:cutoff}
 
In section~\ref{Sec:IR} we presented the numerical computation of the variance by considering the perturbation scale from $k_\text{IR}$ to $k_\text{UV}$.\footnote{In this study, we set $k_\text{UV}=300\mathcal H_o\simeq 0.1 h/\text{Mpc}$ since the nonlinear effect in the matter power spectrum becomes important for $k\gtrsim 0.1 h/\text{Mpc}$~\cite{JEKO06,SMSHSC08}. The detailed study of $k_\text{UV}$ dependence on the luminosity distance variance is presented in \cite{BEGAET12a}.} In this section we present a thorough investigation of the long-mode perturbations at $k \leq k_\text{IR}$. 

For given scale $\bar r$ between the observer and the source, we set $k_\text{IR}$ to be $k_\text{IR}\bar r\ll 1$, and the long-mode gravitational potentials $\Psi_\ell$ can be expressed with the curvature perturbation $\zeta$ as

\begin{align}
\label{Eq:Psi_long}
	\Psi_\ell(\tau,\bar r \hat {\bm n}) =D_\Psi(\tau)\zeta_\ell(\bar r \hat {\bm n})=&~D_\Psi(\tau) \int_0^{k_\text{IR}} \frac{k^2dk}{4\pi^2}\int_{-1}^1 {d\mu}{}~\zeta_{\bm k }~e^{i k \bar r\cos(\theta)} \nonumber\\
	=&~D_\Psi(\tau) \left(\zeta_o + \bar r\hat n^i \partial_i\zeta|_o+\frac{1}{2}\bar r^2 \hat n^i\hat n^j \partial_i\partial_j\zeta|_o+\cdots\right)\nonumber\\
	 \equiv &~D_\Psi(\tau)\left(\zeta_o+\zeta_1(\mathcal H_o\bar r)+\frac{1}{2}\zeta_2(\mathcal H_o\bar r)^2+\cdots \right)\, ,
\end{align}
where $\theta$ is the angle between the wave vector $\bm k $ and the unit vector $\hat {\bm n}$. Dimensionless $\zeta_o$, $\zeta_1$, and $\zeta_2$ are defined as $\zeta_o=\zeta|_o$, $\zeta_1 \equiv \mathcal H_o^{-1}\hat n^i \partial_i\zeta |_o$, and $\zeta_2 \equiv  \mathcal H_o^{-2}\hat n^i\hat n^j \partial_i\partial_j\zeta|_o$, respectively. The suffix $o$ indicates that a quantity is evaluated at the observation i.e., $\bar r=0$.  

According to the equivalence principle, the uniform gravitational potential $\zeta_o$, and the gravity $\zeta_1$ modes should have no effect on local observables.  In appendix \ref{App:EP}, it is shown that the linear-order fluctuation in the luminosity distance  is completely devoid of the contributions of $\zeta_o$ and $\zeta_1$, considering \emph{all terms} in Eq. (\ref{Eq:components}). Thus, the dominant long-mode contribution to $\delta\mathcal D_L$ manifests with $\zeta_2$, and the variance of the dominant long-mode contribution is proportional to $\left<\zeta_2^2\right>$. The variance of $\zeta_2$ is
\begin{equation}
	\left<\zeta_2^2\right> = \frac{1}{20\mathcal H_o^4} \int_0^{k_\text{IR}}  k^3\Delta_\zeta^2(k)~dk\, ,
\end{equation}
\begin{figure}[t]
\centering
\includegraphics[scale=0.7]{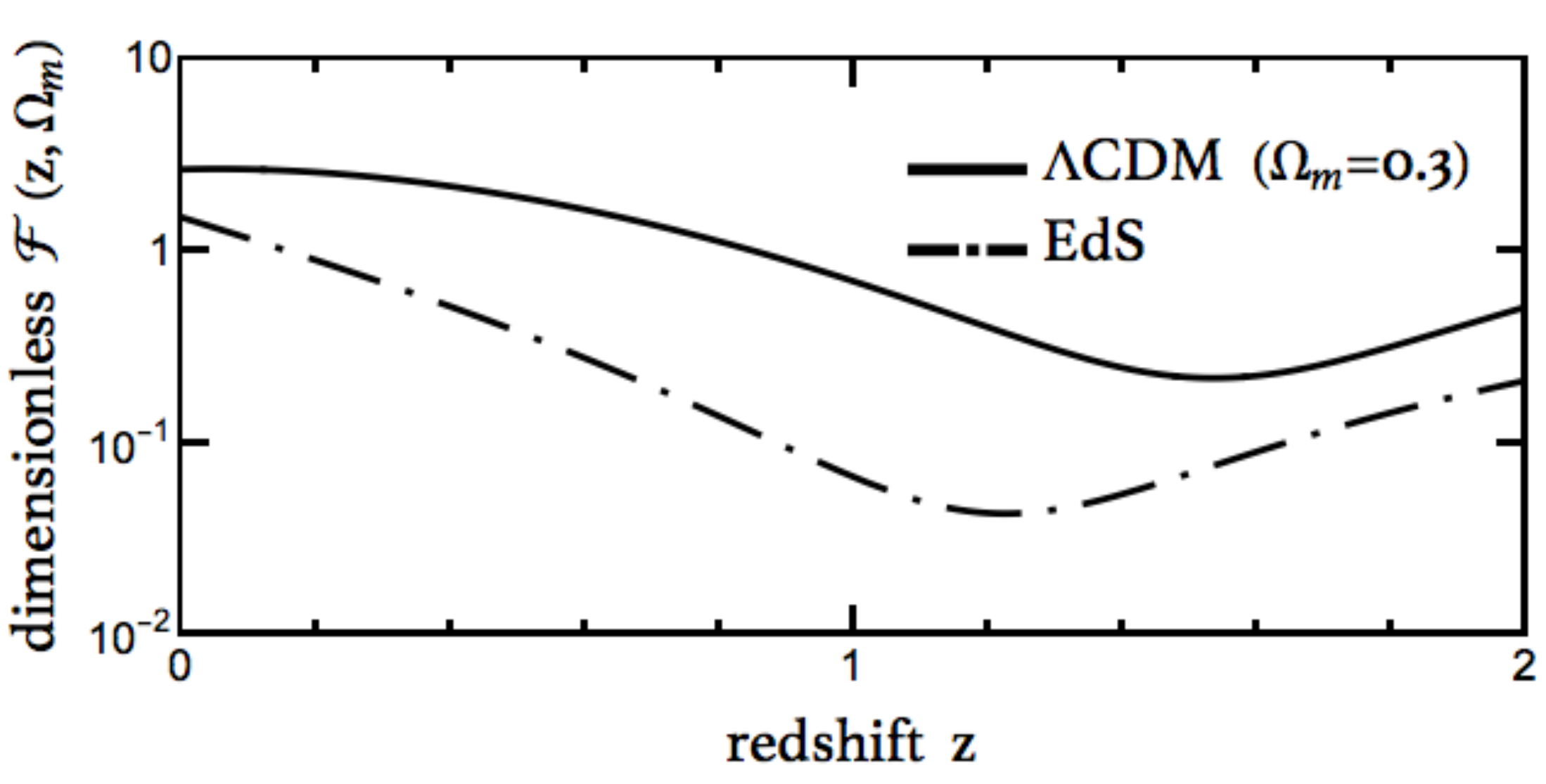}
\caption{The numerical calculation of $\mathcal F$, which contains the late-time universe information in the luminosity distance variance of the long-mode contribution. The solid and dot-dashed lines indicate the cases of the flat $\Lambda$CDM ($\Omega_m=0.3$) Universe and the Einstein-de Sitter (EdS) Universe, respectively. $\mathcal F$ is finite ($\mathcal F_{\Lambda \text{CDM}}\simeq 2.6$ and $\mathcal F_\text{EdS}\simeq1.5$) at $z=0$. It decreases (increases) monotonically for $z\lesssim 1.6$ ($z\lesssim 1.6$) for the $\Lambda$CDM universe ($\Omega_m=0.3$).}
\label{Fig:Ftot}
\end{figure}
where $\Delta_\zeta^2$ is the the dimensionless power spectrum of the curvature perturbation. In the case of $k_\text{IR}\ll k_\text{eq}$, $\left<\zeta_2^2\right>$ is expressed as  $\left<\zeta_2^2\right>\propto (k_\text{IR}/\mathcal H_o)^{4}\left({k_\text{IR}}/{k_o}\right)^{n_s-1}$. Considering a flat $\Lambda$CDM universe, the explicit expression of the dominant long-mode contribution to the variance of the luminosity distances can be written as 
\begin{equation}
\label{Eq:longmode}
	\left<\delta\mathcal D_L^2(z,\hat n)\right>_\ell 
	=\left(\frac{k_\text{IR}}{\mathcal H_o}\right)^4\left\{\hat A_s  \left(\frac{k_\text{IR}}{k_o}\right)^{n_s-1}\mathcal F(z,\Omega_m)\right\}  \, ,
\end{equation}
where $\hat A_s\equiv \frac{8A_s}{375(3+n_s)} =\mathcal O(10^{-11})$ for the best-fit parameters of the primordial universe ($A_s=2\times 10^{-9}$ and $n_s=0.96$), $\Omega_m$ is the current matter amount, and $z$ is the source's redshift. The first two terms $\hat A_s$ and $(k_\text{IR}/k_o)^{n_s-1}$ in bracket are related to the primordial universe, and the last term $\mathcal F(z,\Omega_m)$ contains the information of the late-time universe. The explicit form of $\mathcal F$ is 
\begin{align}
\label{Eq:Ftot}
	\mathcal F(z,\Omega_m) =&~\Bigg[\frac{7}{72}C^2\bar r^4_z-\frac{1}{6}(3D_o-D_z)C\bar r^2_z+\frac{C\bar r_z}{3\mathcal H_z}\left\{2D_o-2D_z+\left(\mathcal  H_z\int_0^{\bar r_z} Dd\bar r\right)\right\} \\
	+&\left\{D_o^2+D_oD_z+\frac{3}{2}D_z^2\right\}-\frac{1}{\mathcal H_z\bar r_z}\left\{(D_o-2D_z)(D_o-D_z)+(3D_o+4D_z)\left(\mathcal H_z\int_0^{\bar r_z}D d\bar r\right)	\right\}\nonumber\\
	+&\frac{1}{2\mathcal H_z^2\bar r_z^2}\left\{3(D_o-D_z)^2-2(D_o-D_z)\left(\mathcal H_z\int_0^{\bar r_z}D d\bar r\right)+7\left(\mathcal H_z\int_0^{\bar r_z}D d\bar r\right)^2\right\}\Bigg]\frac{1}{\Omega_m^2}\, ,\nonumber
\end{align} 
where $D_o$ ($D_z$) means the growth function of the matter field at today (redshift $z$), and $\Omega_{m}$ is the matter amount at today. $C$ is defined as $C=-\frac{5}{2}\mathcal H_o^2\Omega_m$ (see also Eq. (\ref{eq:solution})).

As shown in appendix~\ref{App:EP}, the velocity, the gravitational potential, and the lensing contributions depend on the $\zeta_2$ mode, and $C$ in Eq. (\ref{Eq:Ftot}) represents the gravitational potential contribution. Thus, in the case of the long-mode contribution to the luminosity distance variance, all contributions are important.

Figure \ref{Fig:Ftot} illustrates the numerical calculation of $\mathcal F$ in the flat $\Lambda$CDM universe ($\Omega_m=0.3$) and the Einstein-de Sitter Universe with respect to $z$.  The numerical value of $\mathcal F$ is finite at $z=0$ ($\mathcal F_{\Lambda \text{CDM}}\simeq 2.6$ and $\mathcal F_\text{EdS}\simeq1.5$).\footnote{    In the limit of $z\rightarrow 0$, $\delta\mathcal D_L(z\rightarrow 0,\hat n^i)$ in Eq. (\ref{Eq:dDL}) becomes
\begin{equation}
\label{Eq:DLo}
	\delta\mathcal D_L(z\rightarrow0,\hat n^i) = \frac{3}{2}\mathcal H_o\Omega_m\delta\tau_o + V_{\|o}+\Psi_o+\frac{1}{\mathcal H_o}\left(V_{\|o}'-\partial_\|V_{\|o}+\partial_\|\Psi_o+\Psi_o'\right)\, , 
\end{equation}
and it is constant at $z=0$. Since the luminosity distance in a homogeneous universe is $\bar{\mathcal D}_L(z\rightarrow 0)=0$, the luminosity $\mathcal D_L=\bar{\mathcal D}_L(1+\delta\mathcal D_L)$ in an inhomogeneous universe and the luminosity distance variance $\left<\mathcal D_L^2(z,\hat n^i)\right>$  vanish at $z=0$. In the EdS universe, $\Phi_v$ can be expressed simply as $\Phi_v=-\frac{\eta}{3}\Psi$, and Eq. (\ref{Eq:DLo}) becomes $\delta\mathcal D_L(z\rightarrow0)=\frac{3}{\bar\tau_o}\delta \tau_o +\Psi_o+\frac{\bar\tau_o^2}{6}\partial_\|^2\Psi_o$. This is in agreement with the result in~\cite{SASAK87}.}The amplitude of $\mathcal F$ decreases  for $z\lesssim 1.6$ and increases for $z\gtrsim1.6$. We can see that $\mathcal F<10$ for $z \leqq2$ and $\mathcal F$ depends weakly on the cosmology model. Thus, $\left<\delta\mathcal D_L^2\right>_\ell$ is $\mathcal O(10^{-11})$ for $k_\text{IR}\simeq \mathcal H_o$. As we will see in the next section, the  amplitude of the contribution of the sub-horizon perturbations to the luminosity distance variance is much larger than $\mathcal O(10^{-11})$, and the result in this section implies that the contribution of the super-horizon perturbations to the luminosity distance variance  is \emph{negligible}. In the following section, we study the contribution of the sub-horizon perturbations to the luminosity distance variance.

\section{Variances of the luminosity distance}
\label{Sec:var}

\begin{figure}[t]
\centering
\includegraphics[scale=0.8]{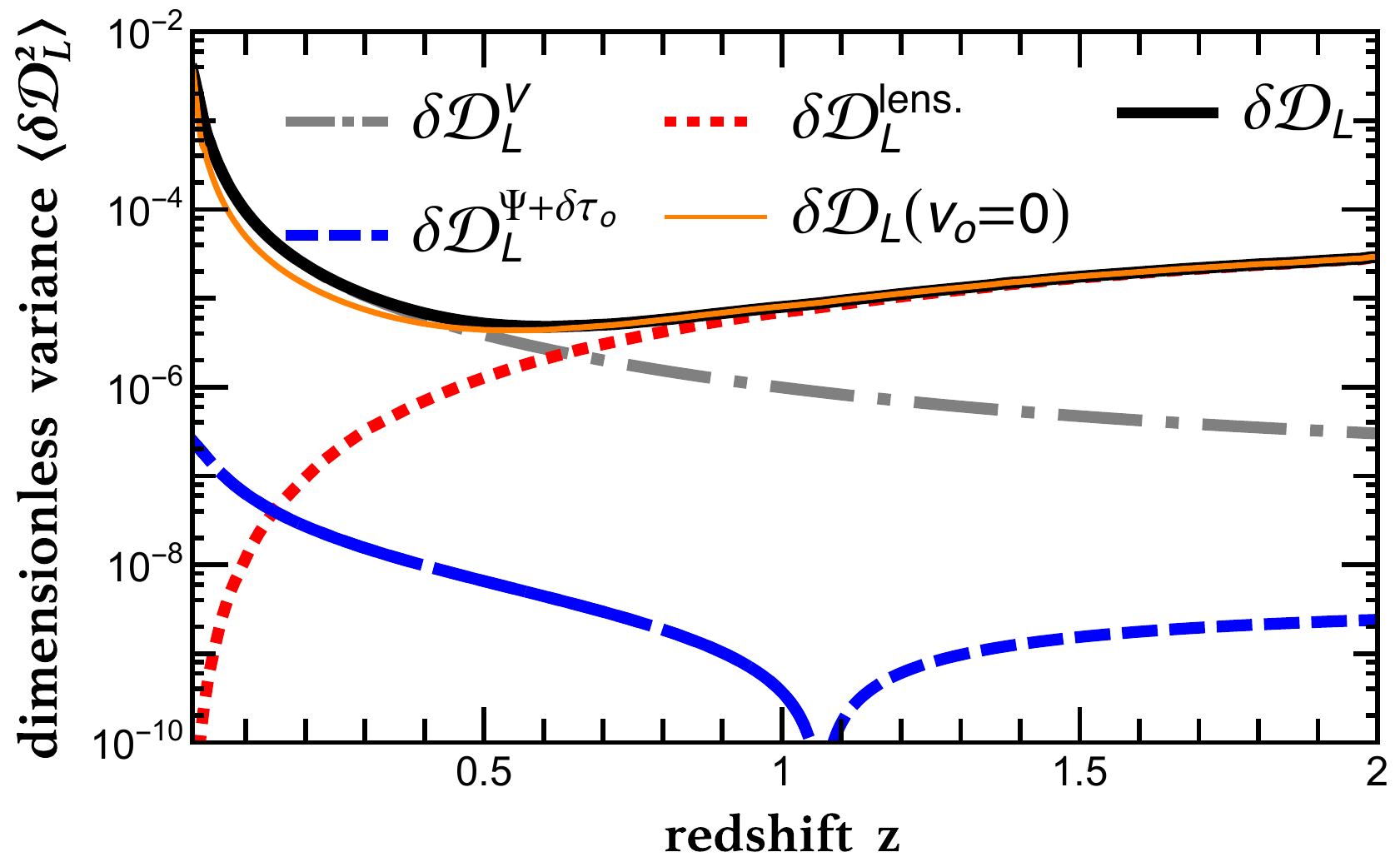}
\caption{Contributions to the variance $\left\langle\ddL^2\right\rangle$ 
of the luminosity distance at each redshift.  The dot-dashed gray and the dotted red lines indicate respectively the velocity $\delta D_L^V$ and the lensing contributions $\delta D_L^\text{lens.}$. The variance is dominated by 
the velocity contribution at low redshift ($z\lesssim 0.6$) and the lensing 
contribution at high redshift ($z\gtrsim 0.6$). The dashed blue line corresponds to the combined contribution of the gravitational potential and the coordinate lapse at the observation, and it is
always subdominant in all time at any scales. The thick black line $\delta\mathcal D_L$ shows the variance of all contribution. The thin orange curve  $\delta\mathcal D_L(V_o=0)$ depicts the variance when all contributions of the observer velocity are ignored. 
}
\label{comp}
\end{figure}

In sections~\ref{Sec:IR} and~\ref{Sec:cutoff}, we have seen the super-horizon perturbation contribution to the luminosity distance is ineffective, if the correct expression $\delta\mathcal D_L$ is used. In this section we investigate the sub-horizon perturbation contributions in Eq.~(\ref{Eq:components})  to the luminosity distance variance.

By applying the perturbation solutions  in appendix~\ref{App:sol} into Eq. (\ref{Eq:components}), we can derive the variance of the luminosity distance. Figure~\ref{comp} shows the contributions of each component to the
variance of the luminosity distance, all of which are computed in a way
long modes of gravity are absent. 

The dominant contributions to the luminosity distance arise from the velocity contribution $\delta\mathcal D_L^V$ (the dot-dashed grey line) at low redshift ($z\lesssim 0.6$), while the lensing contribution~$\delta\mathcal D_L^\text{lens.}$ (the dotted red line) dominates at high redshift ($z\simeq 0.6$): the velocity (lensing) contribution increases (decreases) as redshift decreases (increases). The reason for this opposite trend originates from the fact that the dominant contribution of $\delta\mathcal D_L^V$ comes from $\delta r/\bar r_z$. By contrast, the lensing contribution $\delta\mathcal D_L^\text{lens.}$ is cumulative, and it becomes significant at high redshift. The combined contribution of the gravitational potential and the coordinate lapse at the observation ~$\delta\mathcal D_L^{\Psi+\delta\tau_o}$ consistently deceases as redshift increases, and it has a sign flip from positive to negative at $z\simeq 1.1$. However, this contribution is never important on all scales as long as the divergence is properly regulated, and its contribution order to the total $\delta\mathcal D_L$ (the solid black line) is $\mathcal O(10^{-4})$.

The observer velocity contribution is commonly neglected in many studies~(e.g., \cite{BEDUET14,HUGR06}), and the total variance without the observer velocity is depicted as the thin solid orange line in Fig.~\ref{comp}.  Turning off the observer velocity by hand leads to the \emph{underestimated} variance of the luminosity distance compared to the variance of all contribuions. In particular, significant errors ($\sim40-50\%$) at $z=0.1-0.5$ are made in the variance calculations. 

Figure \ref{Fig:Ftot} shows the the contribution of the super-horizon perturbation to the luminosity distance variance, and the minimum amplitude point appears at $z\simeq 1.6$, while the minimum amplitude point manifests at $z\simeq 0.5$ in Fig. \ref{comp}, and their behaviors are different. The reason is that the gravitational potential contribution plays a significant role in the case of the super-horizon perturbation contribution.  When we consider only the super-horizon perturbation contribution to the luminosity distance variance, the magnitudes of all contributions $\delta\mathcal D_L^{V+\text{lens.}+\Psi+\delta\tau_o}$ are comparable. In contrast to the case of the super-horizon perturbation consideration, the gravitational potential contribution is completely negligible in the consideration of the sub-horizon perturbation only. As a result, the shapes of Fig. \ref{Fig:Ftot} and Fig. \ref{comp} are different. In addition, the amplitude of the super-horizon perturbation contribution $\mathcal O(10^{-11})$ is completely suppressed to that of sub-horizon perturbation contribution (larger than $\mathcal O(10^{-6})$).

\section{Discussion}
\label{Sec:discussion}
 
We have computed the variance of the luminosity distance. As opposed to the
cases in literature, where the variance diverges illustrated in Fig.~\ref{IR}
or an {\it ad hoc} cutoff scale~$\kcut$ is imposed by hand, we have demonstrated 
that such infrared divergences are {\it absent} in our gauge-invariant 
calculation $\delta\mathcal D_L$ and its presence (or the need to impose~$\kcut$)
in previous studies owes to the fact that their calculations were based 
on (incorrect) gauge-dependent expressions $\delta\mathcal D_L^\text{std}$. The equivalence principle guarantees that the monopole modes of gravity,
responsible for infrared divergences, cannot affect the local dynamics,
and such monopole modes are indeed cancelled in our calculations.\footnote{Entropy perturbations on super-horizon scales may affect the local dynamics. However, the equivalence principle dictates that such correlations should originate from non-gravitational interactions.}

Despite the absence of the infrared divergence, the super-horizon contribution to the luminosity distance correlation might be substantial since the gravitational potential is proportional to $(\mathcal H_o/k)^2$. Thus, we computed the variance of this contribution, and we found that the contribution of the super-horizon perturbation is negligible, if the correct expression is used.  However, if one inadvertently misses the monopole terms e.g., the coordinate lapse at the observation or the gravitational potential at the observation which are commonly neglected in many studies, this causes the violation of the equivalence principle and breaks gauge-invariance in the luminosity distance. In this case, the super-horizon mode becomes artificially important. However, there is a very subtle numerical difference between the correct result and incorrect one, when we impose a cutoff scale $k_\text{IR}\simeq \mathcal H_o$ by hand.

Fortunately though, the level of the systematic errors is rather small
as long as an {\it ad hoc} infrared cut-off scale~$\kcut$ larger than the
horizon scale is imposed in the calculations (see Fig.~\ref{IR}),
because the leading correction is suppressed by $(\kcut/\HH_o)^4$ as shown in section \ref{Sec:cutoff}.
The variance due to inhomogeneities is at the level of percents at all
redshifts, mainly due to the velocity contribution at low redshift
and the lensing contribution at high redshift. The amplitudes of the luminosity distance variance computed in this paper can be magnified, if larger $k_\text{UV}$ is adopted. Especially, the largest amplitudes of the velocity and the lensing contributions increase respectively 3 times and 10 times for $k_\text{UV}=\infty$, as shown in~\cite{BEGAET12a}.
 
The features and the expressions of the velocity and the lensing contributions are in agreement with the previous studies. However, the observer velocity contribution is often not considered in~\cite{DAVIS11,MAC16,HUGR06,HUSHSC15,BEDUET14,BEGAET12a,BEMAET12a,BEGAET13}, although the source velocity contribution result is consistent with our result. Neglecting the observer velocity contribution underestimates the luminosity distance variance  as shown in Fig. \ref{comp}, and such underestimation can be seen in~\cite{MAC16}, where the magnitude variances of the theoretical source velocity contribution and the simulation result are presented.\footnote{The magnitude variance is obtained by multiplying $(5/\ln(10))^2$ to the luminosity distance variance.} In particular, significant errors ($\sim40-50\%$) at $z=0.1-0.5$ are made in the variance calculations (the thin solid curve in Fig.~\ref{comp}), if all the observer velocity contributions are ignored.

The lensing variance is shown in~\cite{MAC16,BEGAET12a}, but there is a discrepancy between our result and the result in~\cite{MAC16}. The lensing variance in~\cite{MAC16} becomes comparable to the (source) velocity contribution at $z\simeq 0.4$, whereas this happens at $z\simeq0.6$ in this paper. This discrepancy originates from the choices of cutoff $k_\text{UV}$ and the matter power spectrum $P(k)$: $k_\text{UV}=0.1~h/\text{Mpc}$ ($k_\text{UV}=\infty$) and use of the linear matter power spectrum (the nonlinear matter power spectrum) in our paper (in the previous research). As shown in~\cite{BEGAET12a}, the growing behavior of the lensing variance amplitude is more effective than that of the velocity contribution, as increasing $k_\text{UV}$. Thus, the crossing point between the velocity and the lensing contributions manifests at smaller redshift in~\cite{MAC16}.

The mean of the luminosity distance $\langle\ddL\rangle$
itself is also affected by the presence of
inhomogeneities in our universe, as in our calculation of the variance.
Unfortunately, the investigation of the mean requires going beyond the
linear order calculations in our work and checking its gauge-invariance
of the expression at the second order in perturbations.
Moreover, a proper measure needs to be defined as in \cite{GAMAET11}
(see also \cite{HISE05}) to account for
the fact that the average in observation is different from the ensemble
average in perturbation theory, while we suspect that the second-order
expression of~$\ddL$ itself would be devoid of infrared divergences
for the same physical reason, before any averaging is performed
(the lowest order contribution to the variance in our case is
independent of such measure). Quantifying the correct level of the 
shift in the mean of the luminosity distance due to inhomogeneities
would play an important role in interpreting the luminosity distance
measurements with high precision in future surveys.

\acknowledgments
We acknowledge useful discussions with Camille Bonvin, Robert Brandenberger, Chris Clarkson, Ruth Durrer,  Giuseppe Fanizza,  Jinn-ouk Gong, Donghui Jeong, Roy Maartens, Fabian Schmidt, and Gabriel Veneziano. We acknowledge support by the Swiss National Science Foundation, and
J.Y. is further supported by
a Consolidator Grant of the European Research Council (ERC-2015-CoG grant
680886).

\appendix
\section{Derivation of the wave vector distortion}
\label{App:wavevector}

In this appendix we present the propagation equation for the photon wave vector. In an inhomogeneous universe, we parametrize the metric perturbations with four scalar ($\alpha,\beta,\varphi,\gamma$) and the velocity perturbation of the $i$-th spatial component $u_i$ with one scalar ($v$) as
\begin{equation}
\label{Eq:metric}
	ds^2 = a(\tau)^2\left\{ -(1+2\alpha)d\tau^2-2\beta,_id\tau dx^i+\left((1+2\varphi)\delta_{ij}+2\gamma,_{ij}\right)dx^idx^j\right\},~~u_i\equiv av,_i\, ,
\end{equation}
where the comma denotes the spatial partial derivative e.g., $\beta,_i = \partial_i\beta$, and the indices with the comma are  raised/lowered by $\delta _{ij}$ e.g., $\beta,_i=\beta^{,i}$. Note that we define the velocity perturbation with the sub-index $i$. In this case, at the linear order, the spatial component of the contra-variant velocity vector  is $u^i =g^{i\mu}u_\mu= \frac{1}{a} \left(v^{,i}+\beta^{,i}\right)$. \footnote{The indices of four vectors are raised or lowered by $g_{\mu\nu}$. It is also noted that there exist other ways in literature to define the velocity perturbation. } 

Let $\bar k^\mu=(-1,\hat n^i)$ and $\delta k^\mu=(\delta\nu,\delta n^i)$ be the (conformally transformed) unperturbed and the perturbed photon wave vectors. The linear-order solution for $\delta k^\mu$ is derived by integrating the geodesic equation as 
\begin{equation}
\label{Eq:geodesic}
\delta k^\mu(\chi) = \delta k_o^\mu- \int_0^{\chi} d\bar r~\delta(\Gamma^\mu_{\nu\rho}k^\nu k^\rho)  \, ,
\end{equation}
where $\chi$ is a past-directed affine parameter, and the integration is performed along the line-of-sight path $d\bar r$ at the linear order.

With the metric in Eq. (\ref{Eq:metric}), one can derive the $\delta (\Gamma^\mu_{\nu\rho}k^\nu k^\rho)$ as
\begin{align}
\label{Eq:delta_gamma}
	\delta(\Gamma^0_{\nu\rho}k^\nu k^\rho)= & -2\partial_\|\alpha+\partial_\|^2\beta+\partial_\|^2\gamma'+\alpha'+\varphi' \\
	=&~\frac{d}{d\bar r}\left( -2\alpha+\beta'+\gamma''+\partial_\|\beta+\partial_\|\gamma'\right)-\mathcal I '\, ,\nonumber\\
	\delta(\Gamma^i_{\nu\rho}k^\nu k^\rho)=&~\hat n^i \left(\partial_\|\alpha-\partial_\|\beta'+\partial_\|\varphi-2\partial_\|^2\gamma'+\partial_\|^3\gamma-2\varphi'\right)+\partial_\perp^i\left(\alpha-\beta'-\varphi\right)  +\partial_\|^2\partial_\perp^i\gamma-2\partial_\|\partial_\perp^i\gamma' \nonumber\\
	=&~\hat n^i \left\{\frac{d}{d\bar r}\left(\alpha-\beta'+\varphi-\gamma''-\partial_\|\gamma'+\partial_\|^2\gamma\right)+\mathcal I'\right\}+\frac{d}{d\bar r}\left(\partial_\|\partial_\perp^i\gamma-\partial_\perp^i\gamma'\right) +\partial_\perp^i\mathcal I\, ,\nonumber
\end{align}
where $\mathcal I\equiv \alpha-\beta'-\varphi-\gamma''$. We defined the gauge-invariant quantity $\mathcal I$. The perpendicular $\partial_\perp^i$  and the parallel $\partial_\|$ derivatives with respect to the line-of-sight are defined as $\partial_\perp^i \equiv (\delta ^{ij}-\hat n^i\hat n^j)\partial_j$ and $\partial_\|\equiv \hat n^i\partial_i$, respectively. Note that we replaced $\partial_\|$ with $\frac{d}{d\bar r}$ by using $\partial_\|=\frac{d}{d\bar r}+\partial_\tau$, and the total line-of-sight derivative $\frac{d}{d\bar r}$  will be extracted out from the integration in Eq. (\ref{Eq:geodesic}). With Eq. (\ref{Eq:delta_gamma}), one can derive the wave vector distortion at the linear order level as
\begin{align}
	\label{Eq:wave_distortion}
	\delta\nu(\chi) =& ~\delta\nu_o +\left[2\alpha-\beta'-\gamma''-\partial_\|\beta-\partial_\|\gamma'\right]_o^\chi + \int_0^{\chi} d\bar r~\mathcal I'\, ,\nonumber\\
	\delta n_\|(\chi)=&~ \delta n_{\|o}+\left[-\alpha+\beta'-\varphi+\gamma''+\partial_\|\gamma'-\partial_\|^2\gamma\right]_o^\chi-\int_0^{\chi}d\bar r~\mathcal I' \, ,\nonumber\\
	\delta n_\perp^i(\chi)=&~\delta n_{\perp o}^i +\left[\partial_\perp^i\gamma'-\partial_\|\partial_\perp^i\gamma\right]^\chi_o -\int_0^{\chi}d\bar r~\bar\partial_\perp^i\mathcal I\, ,
\end{align}
where $[y]^\chi_o$ indicates $[y]^\chi_o\equiv y_\chi-y_o$, and the subscripts $\chi$ and $o$ represent the position at $x_\chi^\mu = ( \tau_o-\chi,\chi\hat n^i)$ and  $x_o^\mu=x_{\chi=0}^\mu$, respectively. $\bar \partial_\perp^i$ in the integration of $\delta n_\perp^i$ is defined as $\bar \partial_\perp^i\equiv (\delta ^{ij}-\hat n^i\hat n^j)\partial/\partial \bar x^j$, where $\bar x^\mu = (\bar\tau_o-\bar r,\bar r\hat n^i)$ represents the coordinate along the line-of-sight. The boundary terms $\delta\nu_o$ and $\delta n_o^i$ are determined as follows.

In the observer's local Lorentz frame, the physical wave vector  is $p_L^a = \omega_o(-1,\hat e^I)$, where $e^I$ is a unit vector tangent to the observed photon direction and $\omega_o$ is the observed energy. To find the relation between the wave vectors in the local Lorentz frame $p_L^a$ and in the global coordinate frame $k_o^\mu$, we consider the tetrad $e_a^\mu$, describing the orthornormal coordinate in the local Lorentz frame:
\begin{equation}
\label{Eq:tetrad}
	e_t^\mu = u^\mu=\frac{1}{a(\tau)} \left(1-\alpha,v^{,i}+\beta^{,i}\right)	, ~~~~~~~~~~~~~~e_I^0=\frac{1}{a(\tau)}v_I,~~~~~~~~~~~~~~e_I^j=\frac{1}{a(\tau)}\left(\delta^j_I-\varphi\delta^j_I-\partial^j\partial_I\gamma\right)\, ,
\end{equation}
where $t$ and $I(=1,2,3)$ indicate the time and the spatial components in the local Lorentz frame. By projecting $p_L^a =  e^a_\mu p^\mu_o=\omega_o(-1,e^I)$, where $p^\mu_o$ is the physical wave vector i.e., $p^\mu_o = \frac{1}{\mathbb C a(\tau_o)^2}k_o^\mu$  and $\mathbb C $ is the proportional constant between the physical and conformal affine parameters~\cite{WALD}, one can obtain the relation, and the spatial part becomes 
\begin{equation}
\label{eq:eInI}
	\omega_o \hat e^I = \frac{1}{\mathbb C a(\tau_o)} \left\{\hat n^I \left(1 +\delta n_{\|o}+\partial_\|v_o+\varphi_o+\partial_\|\beta_o+\partial_\|^2\gamma_o\right) +\delta n_{\perp o}^I + \partial_\perp^I v_o+\partial_\perp^I\beta_o+\partial_\|\partial_\perp^I\gamma_o\right\}\, .
\end{equation}
We set $\hat e^I$ to be $\hat n^I$, and $\delta n_{\perp o}^i$ is derived as
\begin{equation}
\label{eq:delta_no2}
	\delta n_{\perp o}^i = - \partial_\perp^i v_o-\partial_\perp^i \beta_o-\partial_\|\partial_\perp^i\gamma_o\, ,
\end{equation}
and $\delta n_{\|o}$ is determined:
\begin{equation}
\label{eq:delta_no1}
	\delta n_{\|o} = -\partial_\|v_o-\partial_\|\beta_o-\partial_\|^2\gamma_o+\mathbb X_o\, ,
\end{equation}
where $\mathbb X_o$ is the additional radial component in $\delta n_o^i$ i.e., $\delta n_o^i=\mathbb X_o \hat n^i -\partial^i v_o-\partial^i\beta_o-\partial_\|\partial^i\gamma_o$. From Eq. (\ref{eq:eInI}), the relation between $\mathbb X_o$ and $\mathbb C$ is derived as
\begin{equation}
	\mathbb C \omega_o a(\tau_o)= 1+\varphi_o+\mathbb X_o  \, .
\end{equation}
With this relation and the temporal part $p_L^t$, one can derive $\delta\nu_o$ as
\begin{equation}
\label{eq:delta_nu}
	-\omega_o \mathbb C a(\tau_o)= -1+\delta\nu_o-\alpha_o-\partial_\|v_o,~~\rightarrow~~\delta\nu_o = \alpha_o+\partial_\|v_o-\varphi_o-\mathbb X_o\, .
\end{equation}
Note that $\delta n_o$ and $\delta n_{\|o}$ satisfy the null condition: $\delta \nu_o+\delta n_{\|o} = \alpha_o-\partial_\|\beta_o-\varphi_o-\partial_\|^2\gamma_o$.

To reduce one degree of freedom $\mathbb X_o$, the additional normalization is needed. The normalization $\mathbb X_o=\mathcal H_o\delta\tau_o-\varphi_o$ yields $\mathbb C \omega_oa(\bar\tau_o)=1$. The advantage of this normalization is that the expression of the observed redshift becomes simple as $1+z_s = \frac{a(\bar\tau_o)}{a^2_s}k^\mu u_\mu|_s$, and this normalization is applied in~\cite{JESCHI12}. Alternatively, we can choose $\mathbb X_o=-\varphi_o$ to make $\mathbb C \omega_o a(\tau_o)=1$, and it is utilized in~\cite{YOO14a}. In this case, $\mathbb C$ depends on a gauge choice, and its gauge transformation is $\tilde{\mathbb C}=\mathbb C (1-\mathcal H_o\xi_o^0)$ since $\mathbb C \omega_o a(\bar\tau_o) = 1- \mathcal H_o\delta\tau_o$, where $\omega_o$ and $a(\bar\tau_o)$ are gauge-invariant.

However, the physical observable should be independent from this artificial normalization. As we shall see, the luminosity distance fluctuation only depends on $\delta n_{\perp o}^i$, which manifests in the angular distortion $\kappa$. That is, the luminosity distance fluctuation is independent from $\delta\nu_o$ and $\delta n_{\|o}$, and it is free from  $\mathbb C$ and $\mathbb X_o$.

\section{Derivations of the frame, redshift, radial, and angular distortions}
\label{App:distortion}

As shown in Eq. (\ref{dL}), the linear-order fluctuation $\delta\mathcal D_L$ in the luminosity distance consists of the radial $\delta r$, the angular $\kappa$, the redshift $\delta z$, and the frame $\Xi$ distortions. The frame distortion is simply expressed as  $\Xi = \frac{1}{2a^2}(\delta^{ij}-\hat n^i\hat n^j)\delta g_{ij}$. In this appendix, we derive the other components in the luminosity distance. For busy readers, we first present their expressions at the linear order and give the detailed derivations next:
\begin{eqnarray}
\label{Eq:frameD}
\Xi &=& \varphi_s+\frac{1}{2}\Delta_\perp \gamma_s+\frac{1}{\bar r_z}\partial_\|\gamma_s \, ,\\
\label{Eq:redshiftD}
		\delta z&=&~\mathcal H_o\delta\tau_o+\left[-\varphi-\mathcal I+ \mathcal V_\| \right]^s_o -\int_0^{\bar r_z}d\bar r~\mathcal I'\, ,\\
		\label{Eq:radialD}
		\delta r &=&~\delta\tau_o-\frac{1}{\mathcal H_z}\delta z +\left[ -\beta-\gamma'-\partial_\|\gamma	\right]^s_o +\int_0^{\bar r_z}d\bar r~\mathcal I\, ,\\
		\label{Eq:angularD}
		\kappa &=&-\mathcal V_{\|o} +\frac{1}{\bar r_z}\partial_\|\gamma_o +\frac{1}{2}\Delta_\perp \gamma_s-\frac{1}{2}\left[\mathcal I\right]^s_o -\int_0^{\bar r_z}d\bar r \left\{\frac{\bar r}{\bar r_z}\mathcal I'+\frac{1}{2}\frac{\left(\bar r_z-\bar r\right)\bar r}{\bar r_z}\left(\mathcal I''-\bar \Delta\mathcal I\right)
	\right\}\, ,
\end{eqnarray}
where the gauge-invariant quantity $\mathcal V_\|$ is defined as $\mathcal V_\| \equiv \partial_\|\left(v+\beta+\gamma'\right)$.  

The observed redshift is determined from the ratio of the source $\mathbb C^{-1}a^{-2}k^\mu u_\mu |_s$ and the observed $\mathbb C^{-1}a^{-2}k^\mu u_\mu |_o$ energies. With this the observer infers the emission time $\bar\tau_z$ as
\begin{equation}
	1+z_s = \frac{a^{-2}k^\mu u_\mu |_s}{a^{-2}k^\mu u_\mu |_o} =\frac{a(\bar\tau_o)}{a(\bar\tau_z)} 	\, ,
\end{equation}
where $\bar\tau_o$ corresponds to the observer's proper time and $\bar\tau_z=\bar\tau_o-\bar r_z$ is the inferred emission time, where $\bar r_z$ is the comoving distance of redshift $z$.  With the wave vector distortion and the 4-velocity, the ratio is determined as
\begin{align}
 \frac{a^{-2}k^\mu u_\mu |_s}{a^{-2}k^\mu u_\mu |_o}=& ~\frac{a(\tau_o)}{a(\tau_s)} \left(1+\left[-\delta\nu+\alpha+\partial_\|v\right]^s_o\right) \nonumber\\
=& ~\frac{a(\bar\tau_o)}{a(\tau_s)} \left(	1+\mathcal H_o\delta\tau_o+\left[-\varphi-\mathcal I+ \mathcal V_\| \right]^s_o -\int_0^{\bar r_z}d\bar r~\mathcal I'\right)\, ,
\end{align}
where the gauge-invariant quantity $\mathcal V_i$ is defined as $\mathcal V_i \equiv \partial_i (v+\beta+\gamma')$. Note that $a(\tau_o)$ is expanded as $a(\tau_o=\bar\tau_o+\delta\tau_o)=a(\bar\tau_o)(1+\mathcal H_o\delta\tau_o$).

The redshift distortion $\delta z$ is defined to describe how the observed redshift $z_s$ is distorted compared with the redshift due to the pure Hubble expansion $\bar z$ as
\begin{equation}
	1+z_s = (1+\bar z)(1+\delta z)=\frac{a(\bar\tau_o)}{a(\tau_s)}(1+\delta z)\, ,
\end{equation}
and the redshift distortion $\delta z$ is determined as
\begin{equation}
\label{Eq:dz}
	\delta z= \mathcal H_o\delta\tau_o+\left[-\varphi-\mathcal I+ \mathcal V_\| \right]^s_o -\int_0^{\bar r_z}d\bar r~\mathcal I'\, .
\end{equation}

Since the wave vector is the tangent vector of the photon curve i.e., $k^\mu =dx^\mu/d\chi$, the true spacetime position of the source is derived by integrating the wave vector as
\begin{equation}
\label{Eq:xs}
	x_s^\mu = x_o^\mu + \int_0^{\chi_s} k^\mu d\bar r\, ,
\end{equation}
where $\chi_s$ is the affine parameter at the emission i.e., $\chi_s=\bar r_z+\delta\chi$, where $\delta\chi$ is the affine parameter perturbation. The spacetime position inferred from the observed redshift and angle is $\bar x^\mu_z = (\bar\tau_o-\bar r_z,\bar r_z\hat n^i)$. With this, the spacetime distortion is determined by subtracting the true position with the inferred position as $\Delta x^\mu=x_s^\mu - \bar x^\mu_z$, and the temporal $\Delta\tau$ and the radial $\delta r$ distortions at the linear order level are
\begin{align}
	\Delta\tau =&~\tau_o+ \int_0^{\chi_s} k^0 ~d\bar r - (\bar\tau_o-\bar r_z)
	=~\delta\tau_o+ \int_0^{\bar r_z}\delta \nu ~d\bar r-\delta\chi \, ,\nonumber\\
	\delta r =\hat n_i\Delta x^i= &~ \int _0^{\chi_s}\hat n_i k^i ~d\bar r	-\bar r_z 
	=\int _0^{\bar r_z}\delta n_\| ~d\bar r+\delta\chi \nonumber\\
	=&~\delta\tau_o+\int_0^{\bar r_z}\left(\delta\nu+\delta n_\|\right)d\bar r-\Delta\tau\, ,
\end{align}
where $\delta\chi$ is the affine parameter perturbation which describes the difference between the affine parameter of the observed line-of-sight $\bar r_z$ and that of the true light geodesic $\chi_s$ i.e., $\delta\chi = \chi_s-\bar r_z$. Note that in the last line, $\delta\chi$ is replaced by applying the first line.

From the relation between the inferred coordinate time and the observed redshift $1+z_s = a(\bar\tau_o)/a(\bar\tau_z)$ and Eq. (\ref{Eq:dz}), the relation between $\Delta\tau$ and $\delta z$ is derived as
\begin{equation}
	\Delta\tau = \frac{1}{\mathcal H_z }\delta z\, ,
\end{equation}
where $\mathcal H_z = \mathcal H(\bar \tau_o-\bar r_z)$. With this, the radial distortion can be expressed with the wave vector and the redshift distortions as
\begin{equation}
	\delta r =\delta\tau_o+\int _0^{\bar r_z}d\bar r \left(\delta\nu+\delta n_\|\right)-\frac{1}{\mathcal H_z}\delta z\, ,
\end{equation}
and the integrand is derived from the null condition $k^\mu k_\mu =0$ as
\begin{equation}
	\delta\nu+\delta n_\| = \alpha-\partial_\|\beta-\varphi-\partial_\|^2\gamma = \frac{d}{d\chi}\left(-\beta-\gamma'-\partial_\|\gamma\right)+\mathcal I\, .
\end{equation} 
Thus, the radial distortion becomes
\begin{equation}
	\delta r=\delta\tau_o-\frac{1}{\mathcal H_z}\delta z +\left[ -\beta-\gamma'-\partial_\|\gamma	\right]^s_o +\int_0^{\bar r_z}d\bar r~\mathcal I\, .
\end{equation} 

In contrast to the previous ones, the angular distortion depends on the wave vector distortion at the observation ($\delta n_{o\perp}^i$), and $\delta n_{o\perp}^i$ is determined by the alignment between the observed angles in the physical and the conformal spacetimes: $\hat e=\hat n$. As referred in appendix~\ref{App:distortion}, $\delta n_{o\perp}^i=-\partial_\perp^i\mathcal V_o$, and we use it in the following.

The angular distortion is defined as $\kappa = -\frac{1}{2}\partial_{\perp i}\Delta x_\perp^i$. $\Delta x_\perp^i$ is derived from Eq. (\ref{Eq:wave_distortion}), Eq. (\ref{eq:delta_no2}), and Eq. (\ref{Eq:xs}) as
\begin{align}
	\Delta x_\perp ^i =&~ \int_0^{\bar r_z}\delta n_\perp ^i ~d\bar r \nonumber\\
	=&-\partial_\perp^i \mathcal V_o \bar r_z+\int_0^{\bar r_z}d\bar r\left(-\frac{d}{d\bar r}\partial_\perp^i\gamma -\int_0^{\bar r}d\tilde r~\tilde\partial_\perp^i\mathcal I\right) \nonumber\\
	=&-\partial_\perp^i\mathcal V_o \bar r_z -\left[\partial_\perp^i\gamma\right]^s_o -\int_0^{\bar r_z}d\bar r \int_0^{\bar r} d\tilde r~\tilde \partial_\perp^i\mathcal I \nonumber\\
	=&-\partial_\perp^i\mathcal V_o \bar r_z -\left[\partial_\perp^i\gamma\right]^s_o -\int_0^{\bar r_z} d\bar r~ (\bar r_z-\bar r)\bar\partial_\perp^i\mathcal I\, ,
\end{align}
and the angular distortion is
\begin{align}
\label{Eq:angular}
	\kappa =& -\frac{1}{2}\partial_{\perp i}\Delta x_\perp^i \nonumber\\
	=&-\mathcal V_{\|o}+\frac{1}{\bar r_z}\partial_\|\gamma_o +\frac{1}{2}\Delta_\perp \gamma_s +\frac{1}{2}\int_0^{\bar r_z}d\bar r\frac{(\bar r_z-\bar r)\bar r}{\bar r_z}\bar\Delta_\perp\mathcal I \nonumber\\
	=& -\mathcal V_{\|o} +\frac{1}{\bar r_z}\partial_\|\gamma_o +\frac{1}{2}\Delta_\perp \gamma_s-\frac{1}{2}\left[\mathcal I\right]^s_o -\int_0^{\bar r_z}d\bar r \left\{\frac{\bar r}{\bar r_z}\mathcal I'+\frac{1}{2}\frac{\left(\bar r_z-\bar r\right)\bar r}{\bar r_z}\left(\mathcal I''-\bar \Delta\mathcal I\right)
	\right\}\, ,
\end{align}
where $\Delta _\perp \equiv \partial_{\perp i}\partial_\perp^i $. The relation $\partial_{\perp i} = \frac{\bar r}{\bar r_z}\bar\partial_{\perp i}$ is applied in the middle line. In addition, the $\bar\Delta_\perp \mathcal I$ in the middle line is altered as
\begin{align}
	\Delta_\perp\mathcal I =&~  (\delta_{i}^j-\hat n_i \hat n^j)\bar\partial_j \left\{(\delta_{ik}-\hat n^i \hat n^k)\bar\partial_k \mathcal I\right\}  \nonumber\\
	=&~  \bar\Delta \mathcal I-\frac{2}{\bar r}\bar\partial_\|\mathcal I - \bar\partial_\|^2 \mathcal I \nonumber\\
	=&~ \bar\Delta\mathcal I -\mathcal I''-\frac{2}{\bar r}\frac{d}{d\bar r}\mathcal I - \frac{d}{d\bar r}\left(\partial_\|\mathcal I+\mathcal I'\right)\, .
\end{align}
Note that the relation $\bar\partial_i\hat n_j = \frac{1}{\bar r} (\delta_{ij}-\hat n_i\hat n_j)$ is utilized in the first line. After some tedious calculations of the integration by part, one can derive the last line in Eq. (\ref{Eq:angular}).

\section{Perturbation solutions in the  $\Lambda$CDM}
\label{App:sol}
In this appendix we express the perturbation solutions in the conformal Newtonian gauge in terms of the linear growth function $D$ and the curvature perturbation in the comoving gauge $\zeta$.

The equation and the solution of the linear growth function of the density contrast in a $\Lambda$CDM are well known as
\begin{equation}
	\frac{d^2 D}{da^2}+\frac{3}{2a}(2-\Omega_m(a))\frac{dD}{da}-\frac{3}{2a^2}\Omega_m(a) D =0\, ,~~	\rightarrow~~D\left(a\right) = a ~{}_2F_1\left[\frac{1}{3},1,\frac{11}{6},-\frac{a^3}{\Omega_{m}}\left(1-\Omega_{m}\right)\right]\, ,
\end{equation}
where ${}_2F_1$ is the hypergeometric function, and $\Omega_m(a)$ is the matter amount when the scale factor of the universe is $a$.

From the ADM (Arnowitt-Deser-Misner) equations~\cite{ADM,NH:2004} in the comoving gauge ($v=0=\gamma$, and $\varphi=\zeta$), the perturbation quantities can be written in terms of the growing mode of the density contrast $\delta_+$ or the curvature perturbation $\zeta$ as 
\begin{eqnarray}
\label{eq:solution}
\zeta (\bm x)&\equiv & ~  C\Delta^{-1}\delta_+(\bm x)\, ,\nonumber\\
	\beta(a,\bm x)&= & -a\mathcal H(a)\frac{dD}{da}\Delta^{-1}\delta_+(\bm x)= -\frac{a\mathcal H(a)}{  C}\frac{dD}{da}\zeta \equiv D_\beta (a)\zeta	\, ,
	\end{eqnarray}
where $D_\beta\equiv -\frac{a\mathcal H}{  C}\frac{dD}{da}$ and $ C \equiv -\left(\mathcal H(a)D'+\frac{3}{2}\mathcal H^2\Omega_m(a) D\right)$. Since $\zeta$ does not depend on time, one can simplify $  C$ by taking the limit of a Einstein-de Sitter universe ($a\ll1$, $D\rightarrow a$, $\mathcal H\rightarrow \mathcal H_o\sqrt{\Omega_{m}/a}$, and $\Omega_m(a)\rightarrow1$) as $ C=-\frac{5}{2}\mathcal H_o^2\Omega_{m} $.
 
 The perturbation solutions in the conformal Newtonian gauge ($\beta=0=\gamma$, $\alpha=\Psi$, $\varphi = \Phi$, and $v,_i=V,_i$) can be obtained by transforming the solution in the comoving gauge to the conformal Newtonian gauge.  Under the gauge transformation ($\tilde x^\mu = x^\mu +(T,L^{,i})^\mu$), the perturbation quantities transform as
\begin{equation}
\label{Eq:pert_GT}
	\tilde\alpha=\alpha-\frac{1}{a}(aT)',~~~~~~~~\tilde\beta=\beta-T+ L',~~~~~~~~\tilde\gamma=\gamma-L,~~~~~~~~\tilde\varphi=\varphi-\mathcal H T,~~~~~~~~\tilde v=v+T \, ,
	\end{equation}
and the displacement field $(T,L^{,i})^\mu$ is derived as $T=\beta$ and $L=0$. Thus, the perturbation variables in the conformal Newtonian gauge are determined as $\Psi = -\frac{1}{a}(a\beta)'$, $\Psi=\zeta-\mathcal H\beta$, $\Phi_v=\beta$, and $\delta\tau_o=\Phi_{vo}$. Thus, they can eventually be expressed in terms of the curvature perturbation $\zeta$ as
\begin{equation}
\label{Eq:Psi}
	\delta\tau_o =  ~D_{V o}\zeta _o\, ,~~~~~~~~~~~~~~~~~~~~	\Psi(a,\bm x) =  D_\Psi(a)\zeta(\bm x) \, ,\,~~~~~~~~~~~~~~~~~~~~
	V_i(a,\bm x)=  D_V(a)\partial_i\zeta(\bm x)\, ,
\end{equation}
 where $D_\Psi(a) \equiv  \mathcal H(a)D_\beta(a)-1$ and $D_V(a)\equiv D_\beta(a)$.

\section{Computation check of the luminosity distance with the equivalence principle} 
\label{App:EP}

According to the equivalence principle, the uniform gravitational potential $\zeta_o$ and the uniform gravity $\zeta_1$ in Eq. (\ref{Eq:Psi_long}) should have no physical effect on  local observables. In this section we show that the luminosity distance fluctuation in a $\Lambda$CDM universe is consistent with the equivalence principle.\footnote{Similar computation is presented in~\cite{SCJE12} by assuming an Einstein-de Sitter universe, where the gravitational potential is temporally constant.} Substituting $\Psi = D_\Psi (\zeta_o+\zeta_1 (\mathcal H_o\bar r))$ and $V_\| = \mathcal H_oD_V\zeta_1$ into Eqs. ({\ref{Eq:frameD} - \ref{Eq:angularD}), one can investigate the contributions of $\zeta_o$ and $\zeta_1$ to the frame, the redshift, the radial, and the angular distortions of the luminosity distance. 

There are three types of integration in the distortions: (1) $\int_0^{\bar r_z}\Psi' ~d\bar r$ in $\delta z$, (2) $\int_0^{\bar r_z}\Psi ~d\bar r$ in $\delta r$, and (3) $\int_0^{\bar r_z}\frac{(\bar r_z-\bar r)\bar r}{\bar r_z}\bar\Delta_\perp\Psi' ~d\bar r$ in $\kappa$. Expressing the gravitational potential $\Psi$ in terms of the temporally constant curvature perturbation in the comoving gauge $\zeta$, one can simplify these integrations. By using $\Psi$ as $\Psi = -\frac{1}{2}(\beta'+\zeta)$, the first integration (1) can be manipulated as 
\begin{equation}
	\label{Eq:int2}
	(1)~:~\int_0^{\bar r_z}\Psi'  d\bar r = \int_0^{\bar r}d\bar r \left(-\frac{d\Psi }{d\bar r}+\partial_\|\Psi \right) 	
	=-\left[\Psi \right]^s_o +\mathcal H_o\zeta_1 \int_0^{\bar r_z}d\bar r D_\Psi \, .
\end{equation}
Likewise, the second integration becomes
\begin{equation}
\label{Eq:int1}
	(2)~:~\int_0^{\bar r_z}\Psi  d\bar r = \zeta_o \int_0^{\bar r_z}d\bar r D_\Psi +\mathcal H_o\zeta_1\int_0^{\bar r_z}d\bar r D_\Psi \bar r \, .
\end{equation}
Since $\zeta_o$ and $\bar r$ vanish when we take the perpendicular derivative $\bar \partial_{\perp i}$, we only need to compute $\bar\partial_{\perp i}\zeta_1 $ to calculate the third integration:
\begin{equation}
	\bar\partial_{\perp i} \zeta_1 = \bar\partial_{\perp i} \hat n^k \partial_k\zeta|_o = \frac{1}{\bar r}(\delta^k_i-\hat n^k\hat n_i)\partial_k\zeta|_o\,,~~~~~~~~~~~~~~~~~~~~~~\bar\Delta_\perp\zeta_1=-\frac{1}{\bar r^2} \hat n^k\partial_k\zeta|_o=-\frac{2}{\bar r^2}\zeta_1\, .
\end{equation}
Exploiting the second expression, we derive
\begin{align}
(3)~:~	\int_0^{\bar r_z}d\bar r~\frac{(\bar r_z-\bar r)\bar r}{\bar r_z}\bar\Delta _\perp \Psi  =&~ {\mathcal H_o}\int_0^{\bar r_z} d\bar r\frac{(\bar r_z-\bar r)\bar r^2}{\bar r_z} D_\Psi \bar\Delta_\perp \zeta_1 \nonumber\\
=&~\mathcal H_o\left( -2\int_0^{\bar r_z}d\bar r D_\Psi +\frac{2}{\bar r_z}\int_0^{\bar r_z}d\bar r D_\Psi \bar r\right)\zeta_1\, .
\end{align} 
With these, the distortions are finally determined as 
\begin{align}
	\Xi(\zeta_o,\zeta_1)=&-D_{\Psi s}\zeta_o-\mathcal H_oD_{\Psi s}\zeta_1\bar r_z \, ,\\
	\label{Eq:EP_dz}
	\delta z(\zeta_o,\zeta_1)=&~(-\mathcal H_o D_{Vo}+D_{\Psi s}-D_{\Psi o})\zeta_o+ {\mathcal H_o}\left(D_{Vs}-D_{Vo}+D_{\Psi s}\bar r_z-2\int_0^{\bar r_z}d\bar r D_\Psi \right)\zeta_1 \nonumber\\
	=&~(1+D_{\Psi s})\zeta_o+{\mathcal H_o}\left(D_{Vs}-D_{Vo}+D_{\Psi s}\bar r_z-2\int_0^{\bar r_z}d\bar r D_\Psi \right)\zeta_1 
	\, ,\\
	\delta r(\zeta_o,\zeta_1)=&~\left(D_{Vo}-\frac{1}{\mathcal H_z}(1+D_{\Psi s})+2\int_0^{\bar r_z}d\bar r D_\Psi \right)\zeta_o \nonumber\\
	&~~+\mathcal H_o\left\{ \frac{1}{\mathcal H_z} \left(-D_{Vs}+D_{Vo}-D_{\Psi s}\bar r_z+2\int_0^{\bar r_z}d\bar r D_\Psi \right)+ 2\int_0^{\bar r_z}d\bar r D_\Psi \bar r 
	\right\}\zeta_1\, ,\\
	\kappa(\zeta_o,\zeta_1)=&~ \mathcal H_o\left(-D_{Vo} -2\int_0^{\bar r_z}d\bar r D_\Psi +\frac{2}{\bar r_z}\int_0^{\bar r_z}d\bar r D_\Psi \bar r\right)\zeta_1\, .
\end{align}
From above, the contribution of the uniform gravitational potential $\zeta_o$ to  $\delta\mathcal D_L$ is derived as
\begin{equation}
	\delta\mathcal D_L({\zeta_o})= \left(1+\frac{1}{\bar r_z} D_{Vo}-\frac{1}{\mathcal H_z\bar r_z}(1+D_{\Psi s}) +\frac{2}{\bar r_z}\int_0^{\bar r_z}d\bar r D_\Psi \right)\zeta_o\, .
\end{equation}
Combining $\Psi =-\frac{1}{a}(a\beta)'$ and $\beta'=-2\mathcal H\beta+\zeta$ in appendix  \ref{App:sol}, we re-arrange each component as  
\begin{equation}
\label{Eq:integ1}
	D_\Psi = -\frac{1}{2} (D_V'+1)\,,~~~~~~~~~~~~~~~~~~~~~\int_0^{\bar r_z}d\bar r D_\Psi = \frac{1}{2}(D_{Vs}-D_{Vo})-\frac{1}{2}\bar r_z\, ,~~~~~~~~~~~~~~~~~~~~~D_\Psi = \mathcal HD_V-1 \, .
\end{equation}
Finally, we demonstrate that the contribution of $\zeta_o$ is indeed zero:
\begin{equation}
	\delta\mathcal D_L({\zeta_o}) =\left(\frac{1}{\bar r_z}D_{Vs}-\frac{1}{\mathcal H_z\bar r_z}(1+D_{\Psi s})\right)\zeta_o =0\, .
\end{equation}
With the above derivations, one can readily check that $\zeta_1$ contribution to $\delta\mathcal D_L$ also vanishes 
\begin{align}
\delta	\mathcal D_L(\zeta_1)=&~  {\mathcal H_o}\left\{D_{Vs}+\frac{1}{\mathcal H_z\bar r_z} \left(-D_{Vs}+D_{Vo}-D_{\Psi s}\bar r_z+2\int_0^{\bar r_z}d\bar r D_\Psi \right)
\right\}\zeta_1 \nonumber\\
=&~  {\mathcal H_o}\left\{D_{Vs}-\frac{1}{\mathcal H_z}(1+D_{\Psi s})\right\} \zeta_1 
=0\, .
\end{align}
Thus, the luminosity distance fluctuation is devoid of $\zeta_o$ and $\zeta_1$ contributions, and it is consistent with the equivalence principle. However, ignoring the coordinate lapse at the observation or the observer velocity in the luminosity distance yields the equivalence principle violation results as
\begin{align}
	\delta\mathcal D_L^{\text{w.o}~\delta\tau_o}(\zeta_o,\zeta_1)=&-\left(\mathcal H_o^2+\frac{\mathcal H_o}{\bar r_z}-\frac{\mathcal H_o^2}{\mathcal H_z\bar r_z}\right)D_{Vo}\zeta_o \, ,\\
	\delta\mathcal D_L^{\text{w.o}~V_{\|o}}(\zeta_o,\zeta_1)= &~\frac{\mathcal H_o}{\mathcal H_z\bar r_z} D_{Vo}\zeta_1\, ,
\end{align}
where $\text{w.o}$ stands for "without".

\end{document}